\documentclass[preprint]{epl2}
\usepackage[latin1]{inputenc}
\usepackage{graphicx}
\usepackage{amssymb,amsmath}

\title{Period doubling route to chaos in Taylor-Green dynamo}

\author{R. Yadav\inst{1} \and M. Chandra\inst{1}  \and M. K. Verma\inst{1} \and S. Paul\inst{1} \and P. Wahi\inst{2} }
\shortauthor{R. Yadav \etal}
\institute
{ \inst{1} Department of Physics, Indian Institute of Technology, Kanpur, India\\
  \inst{2} Department of Mechanical Engineering, Indian Institute of Technology, Kanpur, India
}


\abstract { We perform spectral simulations of dynamo for magnetic
Prandtl number of one with Taylor-Green forcing. We observe dynamo
transition through a supercritical pitchfork bifurcation.   Beyond the
transition, the numerical simulations reveal complex dynamo states
with windows of constant, periodic, quasiperiodic, and chaotic
magnetic field configurations. For some forcing amplitudes, multiple attractors
were obtained for different initial conditions.  We show that one of
the chaotic windows follows the period-doubling route to chaos.}

\begin{document}
 \maketitle

The phenomenon of spontaneous generation of magnetic fields in
magnetohydrodynamics (MHD) is known as the dynamo effect. This is
believed to be the generating mechanism for the magnetic field in
astrophysical bodies such as planets and
stars~\cite{Moffatt_book-Krause_book}.  Dynamo has been observed in
several numerical simulations and laboratory experiments in which
a magnetic field is amplified significantly from a seed magnetic
field. 

Some of the important problems in dynamo are related to the dynamo
transition, e.g., what is the nature of the bifurcation from the fluid state
to the dynamo state with a magnetic field?,
how is chaos generated in dynamo?, what leads to the reversals of
magnetic field?, how is dynamo affected by forcing and geometry?,
etc. In this letter we attempt to understand the nature of
bifurcations and the routes to chaos in dynamo under the
Taylor-Green (TG) forcing.

The governing equations for dynamo are
\begin{eqnarray}
\partial_{t}\mathbf{v}+ (\mathbf{v} \cdot \nabla) \mathbf{v} & = &
-\nabla p+ (\mathbf{B} \cdot \nabla)\mathbf{B}+\nu\nabla^{2}\mathbf{v}+\mathbf{F}, \label{eq:MHD_vel}\\
\partial_{t}\mathbf{B}+ (\mathbf{v} \cdot \nabla) \mathbf{B} & = &
(\mathbf{B} \cdot \nabla) \mathbf{v}+\eta\nabla^{2}\mathbf{B}, \label{eq:MHD_mag}\\
\nabla \cdot \mathbf{v} & = & 0, \\ \label{eq:div_v_0}
\nabla \cdot \mathbf{B} & = & 0,  \label{eq:div_B_0}
\end{eqnarray}
where $\mathbf{v}$ is the velocity, $\mathbf{B}$ is the magnetic
field, $p$ is the total pressure (thermal+magnetic), 
$\mathbf{F}$ is the external force field, $\nu$ is the kinematic viscosity, and
$\eta$ is the magnetic diffusivity.  Some of the important parameters
that characterise dynamo are the magnetic Prandtl number
$P_M=\nu/\eta$, the Reynolds number $Re=UL/\nu$, and the magnetic
Reynolds number $R_M=UL/\eta$, where $U$  and $L$ are the characteristic
velocity and length of the system respectively. Only two of the above three
parameters are independent ($R_M = Re * P_M$).  Note that the range
of  magnetic Prandtl number $P_M$ observed in nature is extremely
large.  Liquid metals and solar plasma typically have small $P_M$
($10^{-5}$ to $10^{-2}$), while interstellar medium has typically
large $P_M$ (of the order of $10^{14}$).  There are other important
parameters in dynamo, for example, Rossby number, Ekman number,
Rayleigh number, etc.  To abstract and simplify the physics of
dynamo transition, in the present letter we focus only on basic MHD
equations [Eqs.~(\ref{eq:MHD_vel}-\ref{eq:div_B_0})] that have only
two independent parameters, $P_M$ and $Re$.

In MHD systems, forcing is typically applied only to the velocity
field.   A steady magnetic field is generated beyond a certain
critical forcing. The magnetic Reynolds number corresponding to this
forcing is called the critical magnetic Reynolds number ($R_M^c$).
The first bifurcation is generally a pitchfork bifurcation.
Subsequent secondary bifurcations yield more complex configurations
like periodic, quasiperiodic, and chaotic magnetic fields.  In this
letter we will investigate the bifurcations and routes to chaos for
$P_M=1$  under the TG forcing.

Dynamo transition has been studied extensively using experiments,
numerical simulations, and theoretical modelling. The Riga dynamo
experiment~\cite{Riga_expt}, which relies on the concept of the
Ponomarenko dynamo~\cite{Dormy:book}, was the first successful liquid metal experiment
in which dynamo was observed.  Dynamo has also been observed in the
Karlsruhe experiment~\cite{Karlsruhe_expt} in which the flow pattern
is helical.  Recently Monchaux {\em et al.}~\cite{Monchaux_vks}
performed dynamo experiments on liquid sodium ($P_M\sim 10^{-5}$)
confined within a horizontal cylinder. The fluid was forced using
two fans at the two ends of the cylinder.  The speeds of the two
fans were used as the main controlling parameters.  This experiment
is called VKS (von K\'arm\'an Sodium) due to the nature of its
forcing and the velocity configurations. Monchaux {\em et
al.}~\cite{Monchaux_vks} observed dynamo for $R_M \approx 30$ for
equal and opposite forcing frequencies of the two fans as a result
of a supercritical pitchfork bifurcation.  The resulting magnetic
field is axisymmetric and is constant in time. By changing the
relative frequencies of the fans, the VKS experimental team could
obtain a variety of dynamo solutions including periodic,
quasiperiodic, and chaotic magnetic field
configurations~\cite{VKS_variousstates}. They also observed
reversals of the magnetic field.   

Various researchers have simulated dynamo using direct numerical
simulations (DNS) with random, TG, ABC, and Roberts forcing and
observed dynamo in both low and high $P_M$ regimes. Variations of
$R_M^c$ as a function of $P_M$
have been investigated in these studies. TG forcing has certain
similarity with  the von K\'arm\'an flow configuration and the VKS
experiment, hence it has become quite popular.  Nore {\em et al.}~\cite{Nore_letter} demonstrated the existence of dynamo action under TG forcing. Ponty {\em et
al.}~\cite{Ponty_lowP} applied TG forcing and reported that $R_M^c$
increases sharply with $P_M^{-1}$  as turbulence sets in, and then
saturates.  Ponty {\em et al.}~\cite{Ponty_subcritical} observed
subcritical dynamo transition by changing $R_M$ in their simulations.
Mininni~\cite{Mininni_dynamo_regimes_TG}  observed various dynamical
regimes including time-periodic oscillations and well-defined
spatial structures.   Dubrulle {\em et al.}~\cite{bifurcations_TG}
reported various bifurcations in TG flows for both hydrodynamics and
MHD simulations. 

In another set of simulations, Schekochihin {\em et al.}~\cite{Sche_critical_Rm} and Iskakov {\em
et al.}~\cite{Iskakov_fluctuation_dynamo} applied non-helical random
forcing and showed the existence of dynamo for both large $P_M$
and small $P_M$. Podvigina~\cite{Podvigina},
Mininni~\cite{Mininni_turbulent_dynamo_excitation_lowP}, and Mininni
and Montgomery~\cite{Mininni_lowP_Roberts_flow} simulated helical
dynamo using ABC and Roberts forcing.  Podvigina~\cite{Podvigina}
simulated dynamo with ABC forcing and studied various magnetic field
states including chaos. The author related these states with inherent
symmetries of the system.  Gissinger {\em et
al}.~\cite{Gissinger_cowley_thm-reversals} generated dynamo by
axisymmetric forced flow in a spherical domain; they also observed
chaotic magnetic field reversals in the same geometry.  Morin and
Dormy~\cite{Morin_subcritical} studied dynamo in a rotating
spherical shell and observed either supercritical or subcritical dynamo
transition depending on the chosen set of parameter values. Glatzmier and Roberts~\cite{Glatzmier_nature} simulated geodynamo and observed several interesting phenomena including the reversals of the magnetic field.

Dynamo transitions have also been studied using low-dimensional
models that are constructed through heuristic arguments, or from
partial differential equations using Galerkin projections. These
models are amenable to analytical investigations of bifurcations due
to smaller number of modes.  Rikitake model or disk dynamo, which
belongs to the former category,  is a discrete model with conducting
discs and current-carrying wires \cite {Rikitake}.  This model
shows complex dynamical behaviour including constant, periodic,
quasiperiodic, and chaotic magnetic fields.  The model of 
Petr\'{e}lis and coworkers~\cite{petrelis_model} 
is based on amplitude equations and symmetry arguments.
They attempted to understand the origin of various dynamo states and magnetic
field reversals using the saddle-node
bifurcation in their low-dimensional model~\cite{petrelis_model}. 
Models based on
Galerkin projections have been constructed by Donner {\em et
al}.~\cite{Donner}  and Verma {\em et al}.~\cite{mkv_lowD}. The
magnetic field in the six-mode model of Verma {\em et
al}.~\cite{mkv_lowD} is generated by a supercritical pitchfork
bifurcation at the transition point. This model exhibits only a
constant magnetic field (in time), and it does not show
complex dynamical behaviour.  On the contrary, the 152-mode model of
Donner {\em et al}.~\cite{Donner} shows various patterns including
constant, time-periodic, quasiperiodic, and chaotic states for
$P_M=1$. There are other dynamo models based on scale separations,
e.g., $\alpha$-dynamo, $\beta$-dynamo,
etc.~\cite{Moffatt_book-Krause_book}.

\begin{center}
\begin{figure}[t]
\onefigure[height=!, width=14cm]{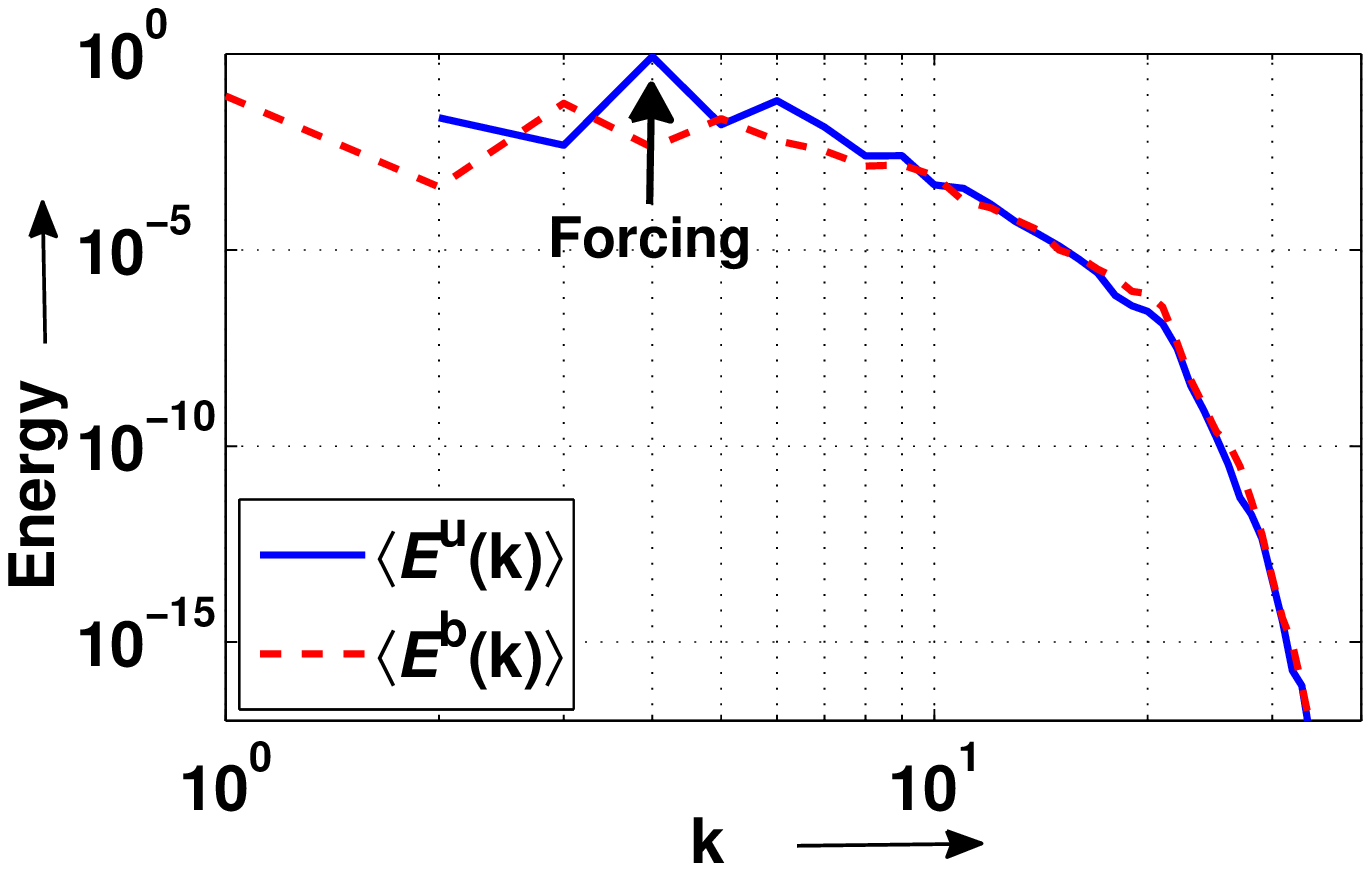}
\caption{} \label{fig:spectrum}
\end{figure}
\end{center}

In this letter we numerically  study the dynamo transition for TG forcing given by
\begin{eqnarray}
\mathbf{F}(k_0) & = & F_0
\left[
\begin{array}{c}
\sin(k_0 x)\cos(k_0 y)\cos(k_0 z) \\
-\cos(k_0 x)\sin(k_0 y)\cos(k_0 z) \\
0
\end{array}
\right], \label{eq:taylor_green}
\end{eqnarray}
where $F_0$ is the forcing amplitude, and $k_0$ is the wavenumber of
the forcing taken as $2$ in this work. Note that the helicity of
the force, $(\nabla \times {\bf F}). {\bf F} = 0$ everywhere.
Nore {\em et al.}~\cite{Nore_letter} and Ponty {\em et al.}~\cite{Ponty_lowP} however argue that local fluctuations in 
kinetic helicity are generated by the above forcing.  We solve
Eqs.~(\ref{eq:MHD_vel}-\ref{eq:div_B_0}) using
TARANG~\cite{Tarang}, a pseudo-spectral code, in a 3D box of
dimensions $2\pi$ on each side for $P_M=1$.  We apply Runge-Kutta fourth order
scheme for time advancement.  The time increment $dt$ is determined
using the CFL condition ($dt = \Delta x/(20 \sqrt{E^u})$, where
$\Delta x$ is the grid size, and $E^u$ is the total kinetic energy).
The number of grid points used in our simulation is $64^3$.
Our runs are dealiased using 2/3 rule. The range of Reynolds number
investivated is from 6 to 160, for which our simulations are well
resolved as $k_{max} \eta$ (the largest wavenumber times
the Kolmogorov length)  is always greater than 1.3.  This
observation is corroborated from the well resolved kinetic energy
spectrum $\langle E^u(k)\rangle$ and magnetic energy spectrum
$\langle E^b(k)\rangle$ shown in fig.~\ref{fig:spectrum} for $\nu =
\eta = 0.1 $, and $F_0=4.8$.
The number of interacting Fourier modes of our dynamo system is
$64^3$, which is quite large.  Our simulations however reveal that
only a small fraction of them carry most  of the energy.  In many
numerical runs with different $F_0$'s, we observe that the most prominent velocity Fourier
modes are $(\pm 2, \pm 2, \pm 2)$,  $(\pm 4, \pm 4, \pm 4)$,  $(\pm
4, \pm 4, 0)$, $(0, \pm 8, \pm 4)$, and the most prominent magnetic
Fourier modes are $(0, 0, \pm 1$), $(0, 0, \pm 2)$, $(0, 0, \pm 3)$,
$(\pm 2, \pm 2, \mp 3)$, $(\mp 2, \mp 2, \pm 1)$. Here the three arguments refer to $x$, $y$, and
$z$ components of the wavenumber. Note that other
modes like $(\pm 4, 0, \pm 4)$ for velocity field are also present
due to symmetry.  The most energetic velocity
Fourier mode is $(\pm 2, \pm 2, \pm 2)$ due to $k_0=2$ of the TG forcing.  Among
the magnetic modes, the most dominant modes are ${\bf B}(0,0,1)$ and ${\bf
B}(0,0,2)$.  The mode ${\bf B}(0,0,1)$ is  generated due to the nonlinear interactions between $({\bf
v}(2,2,2), {\bf B}(-2,-2,-1))$, and the mode ${\bf B}(0,0,2)$ is generated by  $({\bf v}(2,2,2), {\bf B}(-2,-2,0))$.    We observe a
dynamic interplay between  ${\bf B}(0,0,1)$ and ${\bf B}(0,0,2)$ modes.
The other important magnetic mode is ${\bf B}(0,0,3)$ which participates
with $({\bf v}(2,2,2), {\bf B}(-2,-2,-1))$.

\begin{center}
\begin{figure}[t]
\onefigure[height=20cm, width=10cm]{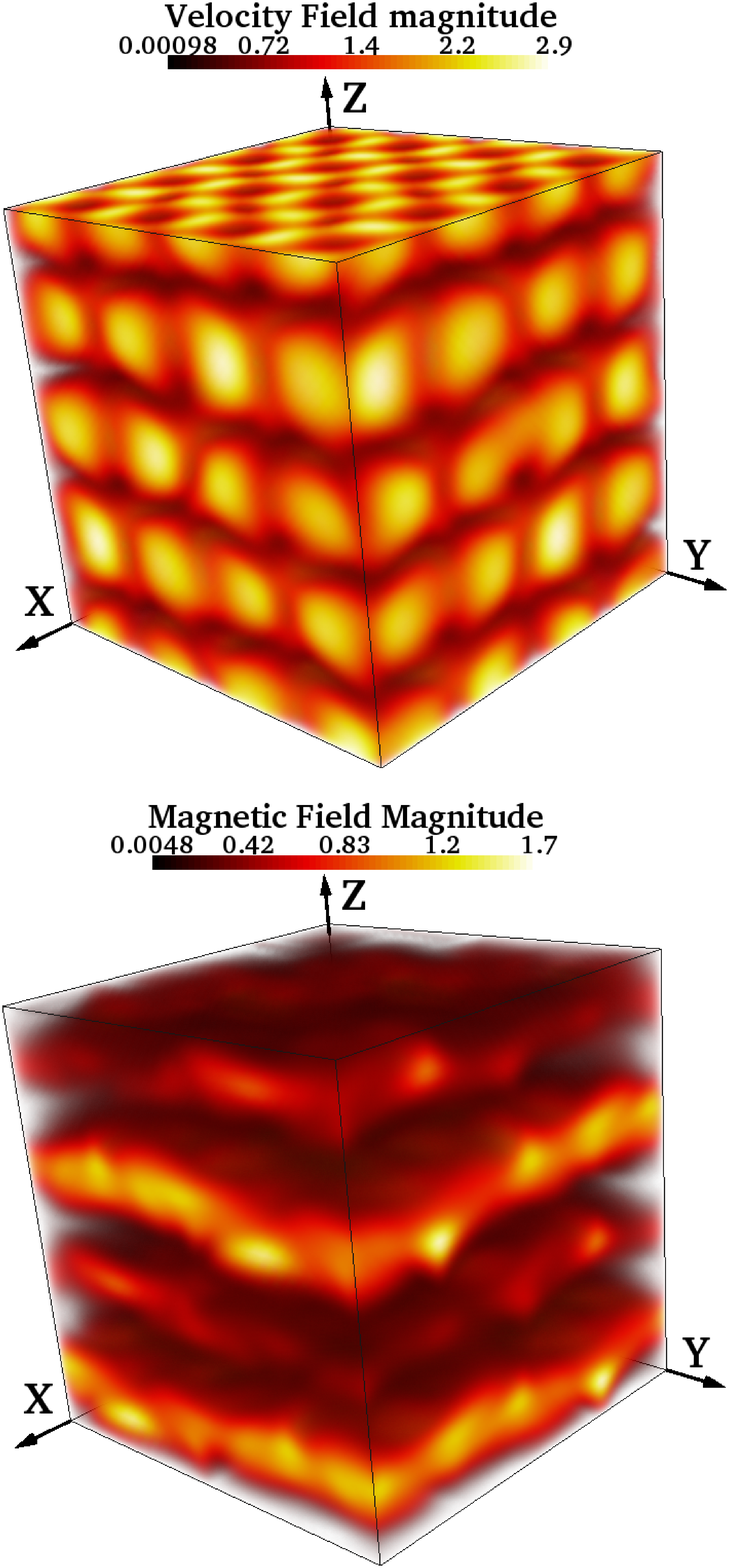}
\caption{} \label{fig:3D}
\end{figure}
\end{center}

\begin{center}
\begin{figure}[t]
\onefigure[height=!, width=14cm]{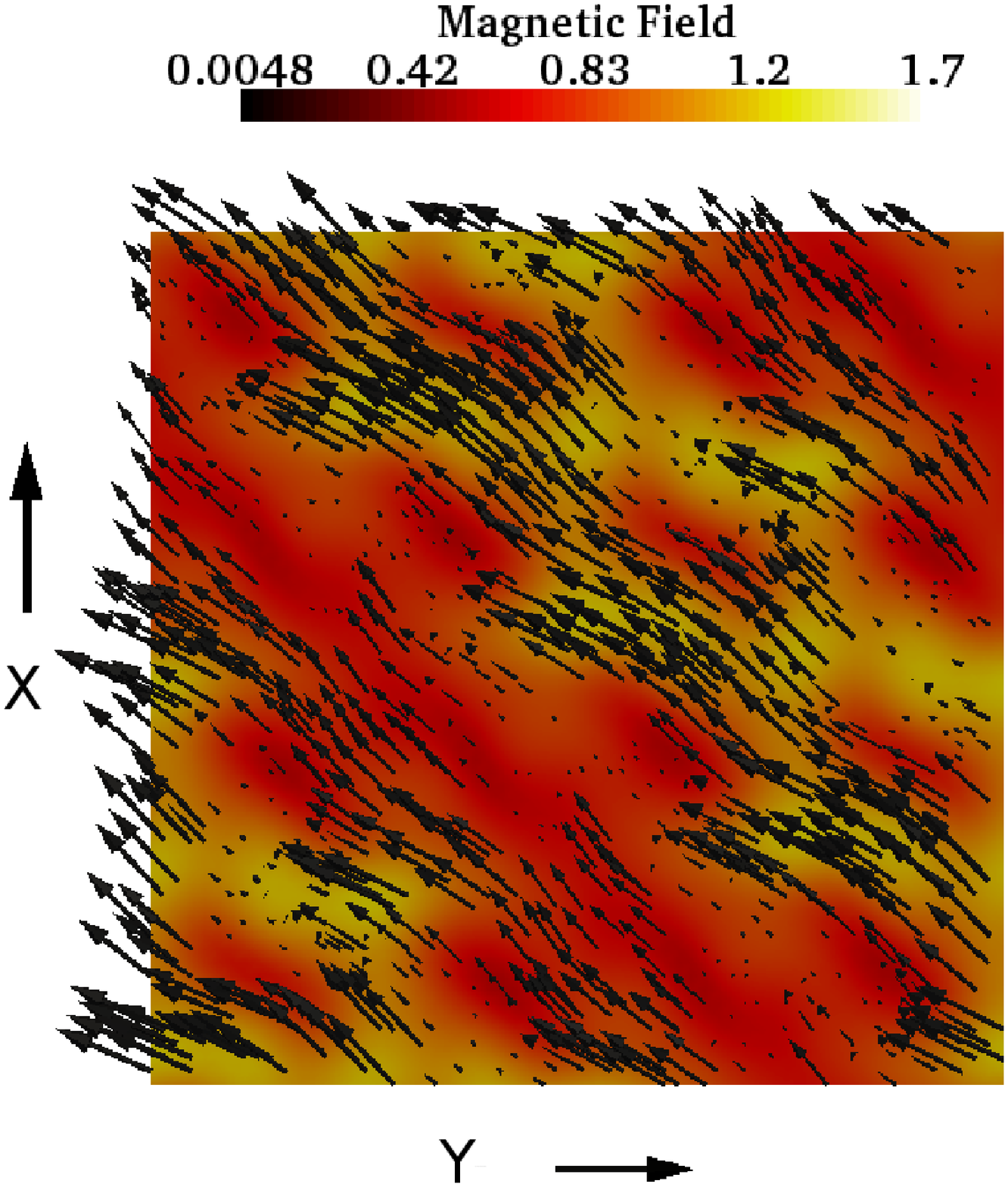}
\caption{} \label{fig:vector}
\end{figure}
\end{center}

We perform DNS for various forcing
parameter $F_0$.  Initial transients are discarded and only the
steady-state configurations are analysed.    For $F_0 = 4.8$, the
steady-state velocity and magnetic fields (snapshots of the
magnitudes) are shown in figs.~\ref{fig:3D}(a,b) respectively. These figures indicate that
the TG forcing yields well defined velocity and magnetic structures.
For $k_0=2$, chosen for our runs, the simulation box has 16 TG cells
for the velocity field.   As shown in fig.~\ref{fig:3D}(b), the magnetic energy is concentrated in two 
major slabs along with two minor slabs.    We also observe that $|B_z| \ll |B_x|, |B_y|$.  The above configuration
is due to  the  prominence of the modes ${\bf B}(0,0,1)$ and  ${\bf B}(0,0,2)$.  Note that the $z$ components for these
modes are zero due to the incompressibility condition. Thus, we can understand the global TG structures in terms of the
dominant Fourier modes.  Figure~\ref{fig:vector} contains the
vector plot of the magnetic field ($x$ and $y$ components only) for
a cross-section of  fig.~\ref{fig:3D}(b) at $z=4.08$.  The
magnetic field in this plane is approximately along $-45^{\circ}$.
These field configurations are similar to  those reported by
Mininni {\em et al}.~\cite{Mininni_dynamo_regimes_TG}.

\begin{center}
\begin{figure}[t]
\onefigure[height=!, width=14cm]{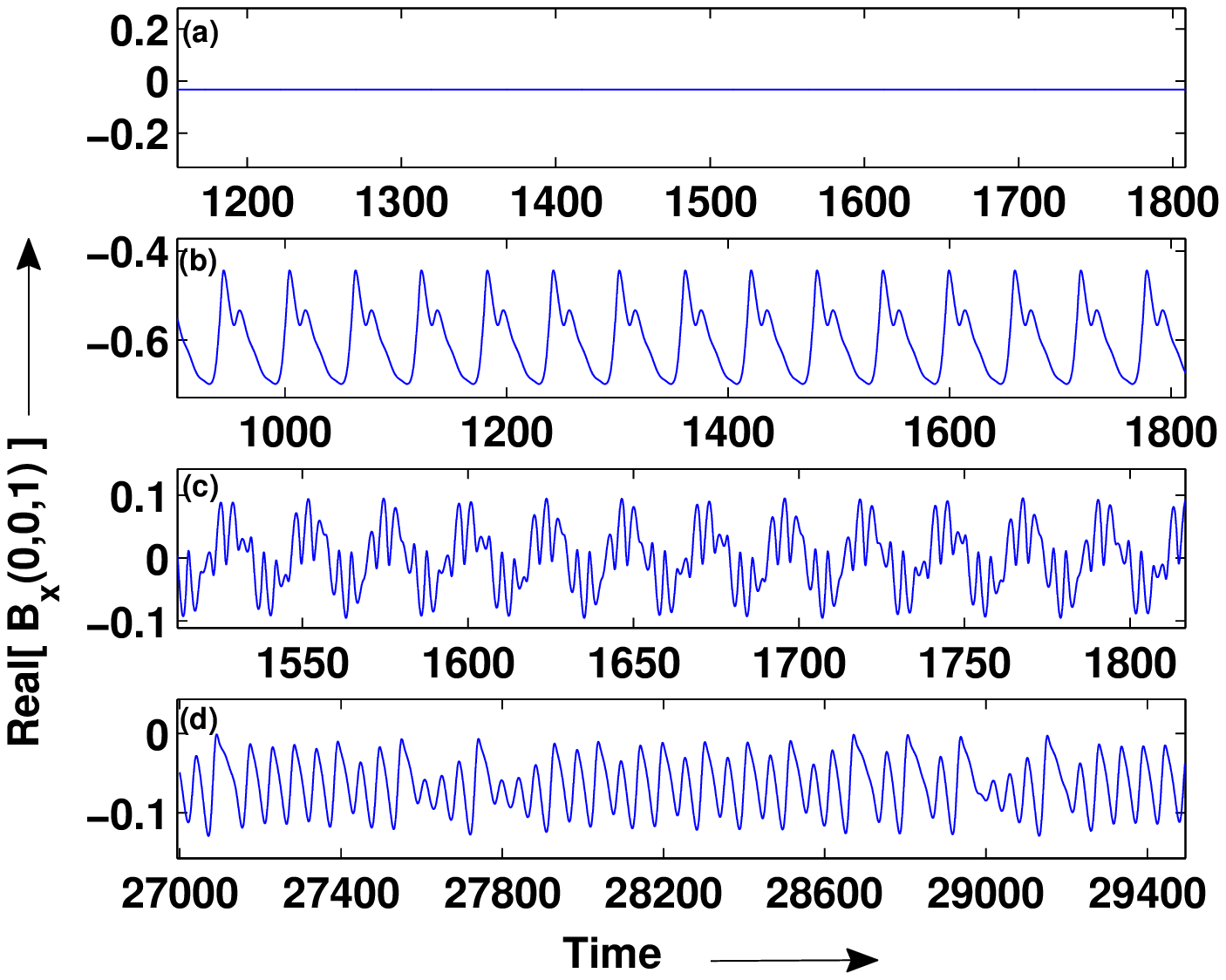} 
\caption{} \label{fig:time_series_states}
\end{figure}
\end{center}

To explore various bifurcations of the dynamo state, we vary $F_0$
in the interval $[1 : 40] $.  We observe pure fluid solutions ($E^b =
0$) till $F_0 = 3.9$, after which nonzero steady-state magnetic
field ($E^b > 0$) emerges with each component of the magnetic field
as constant (in time).  The magnetic Reynolds number at this
transition regime is approximately 19.9, hence $R^c_M \simeq 19.9$.
The number of variables of our dynamo system is unfortunately rather
large ($\sim 64^3$).  We could however focus on the most energetic
modes that determine the system dynamics.  We investigate the time
series of these modes, and obtain various dynamo states: constant (fixed
point), periodic, quasiperiodic, and chaotic magnetic fields.  In
fig.~\ref{fig:time_series_states} we show the time series of a fixed
point ($F_0 = 3.9$), a periodic state ($F_0 = 10$), a quasiperiodic
state ($F_0 = 36$), and a chaotic state ($F_0 = 4.8$) of the dynamo.

\begin{center}
\begin{figure}[t]
\onefigure[height=!, width=14cm]{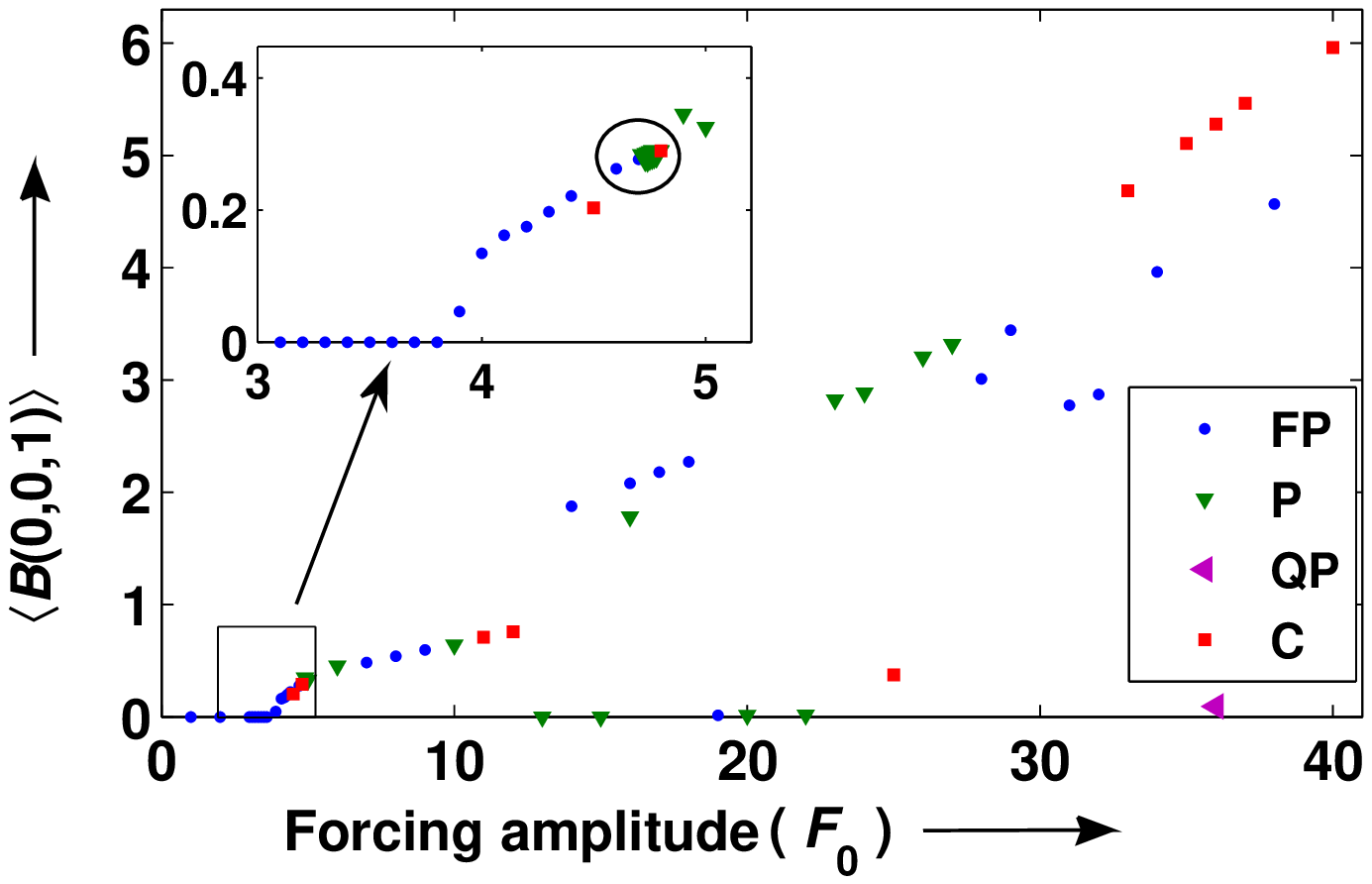}
\caption{} \label{fig:states}
\end{figure}
\end{center}

Different states of a dynamical system are elegantly illustrated in
the bifurcation diagram that contains information about the birth of
new states.  In fig.~\ref{fig:states} we construct a  bifurcation
diagram by plotting the time averaged value of the amplitude of the
magnetic mode $\langle B{(0,0,1)} \rangle $ for different $F_0$
(obtained from around 60 DNS runs).  Nonzero magnetic
field appears at around  $F_0 = 3.9$ through a pitchfork bifurcation.
There is no  hysteresis  near the onset of dynamo as $F_0$ is
increased or decreased.  Hence,  the dynamo transition is through a
supercritical pitchfork bifurcation. After the primary instability (or
bifurcation), we observe different kinds of dynamo states like
constant, periodic,  quasiperiodic, and chaotic
magnetic fields as evident from the bifurcation diagram.  The
windows of these states appear for various range of $F_0$, e.g.,
chaos appears for $F_0 = 11 - 12$.  We also find
windows of $F_0$ where $\langle B{(0,0,1)} \rangle$ becomes
relatively small, but  $\langle B{(0,0,2)} \rangle$ becomes
significant.  The dynamics of interchange of energy between
${\bf B}(0,0,1)$ and  ${\bf B}(0,0,2)$ is not apparent at present, and it
will be studied in future.

\begin{center}
\begin{figure}[t]
\onefigure[height=!, width=14cm]{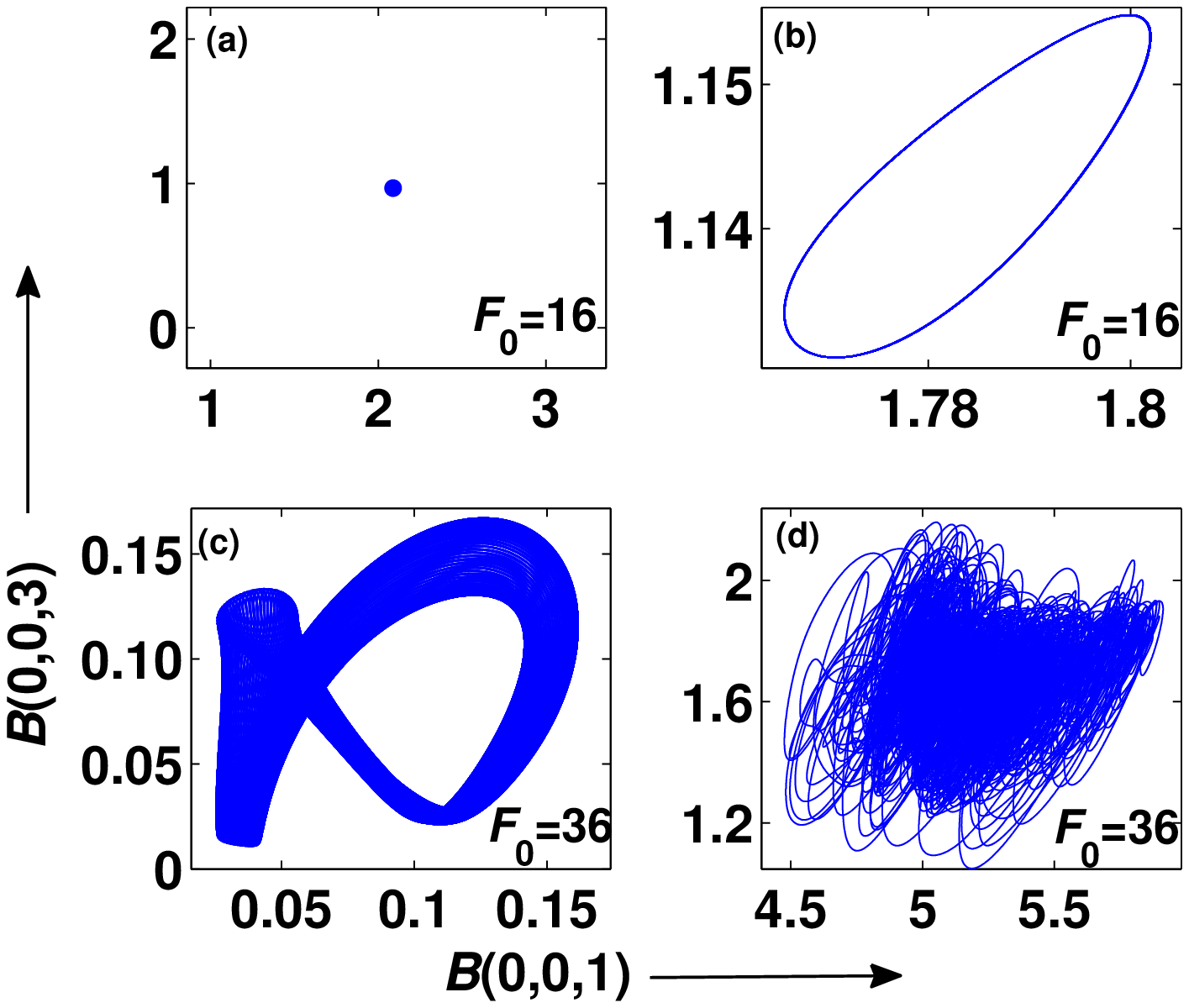}
\caption{} \label{fig:mp}
\end{figure}
\end{center}

A careful analysis of the dynamics for a given forcing reveals
coexistence of multiple dynamo states.  We illustrate this feature
using  state space or phase space.  The phase space of our dynamo is
very large ($\sim 64^3$).  However, projections of the phase space
on the subspace formed by the most energetic modes contain most of the
information of the system.    For $F_0 = 16$, two different sets of
initial conditions yield either a fixed point or a periodic
solution. These two states are illustrated in
figs.~\ref{fig:mp}(a,b) using projections of the phase space on
($B(0,0,1)$-$B(0,0,3)$) plane.  At $F_0 = 36$, the two different coexisting
dynamo states involve quasiperiodic and chaotic magnetic fields as
illustrated in figs.~\ref{fig:mp}(c,d).

\begin{center}
\begin{figure}[t]
\onefigure[height=20cm, width=14cm]{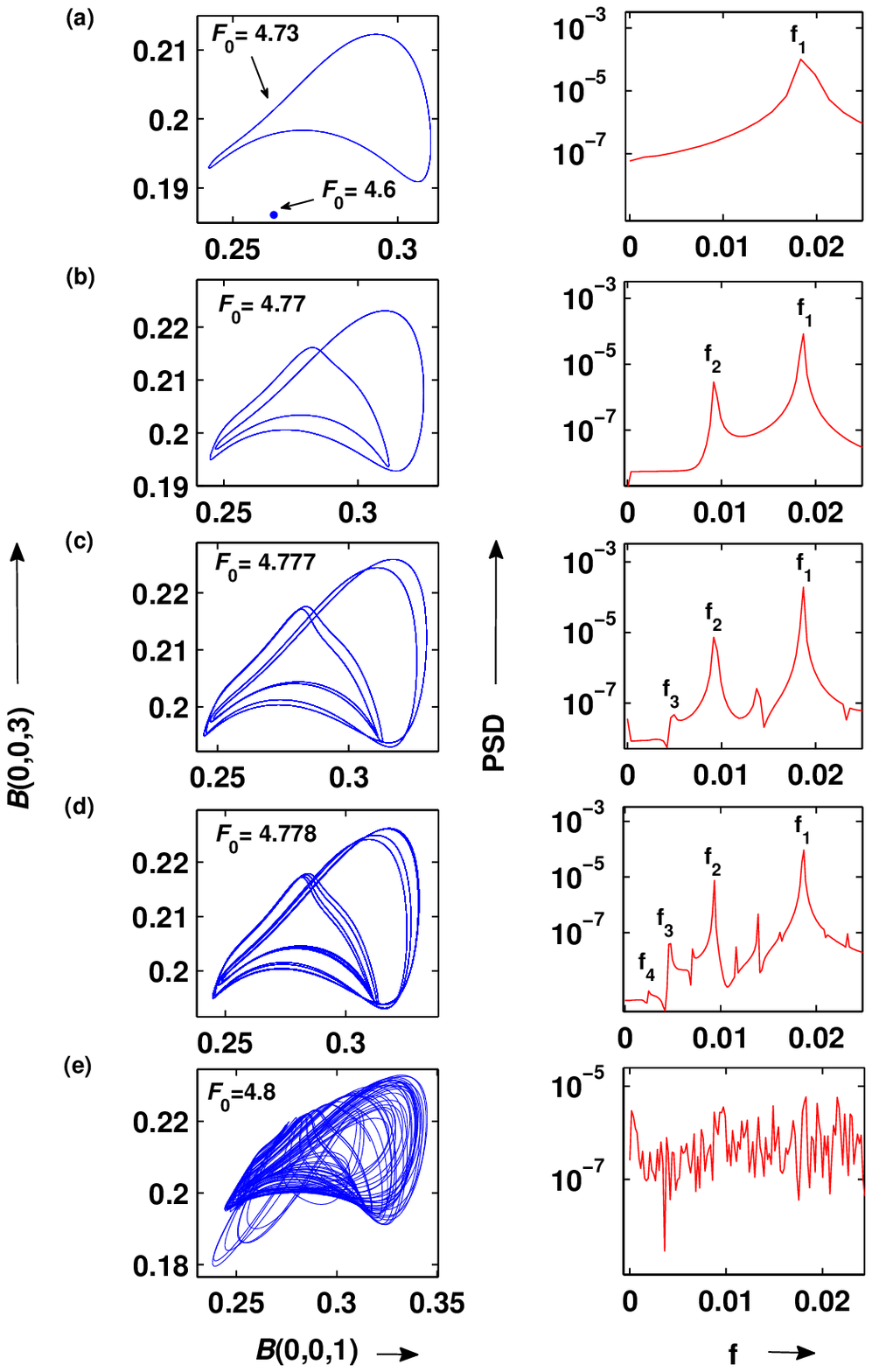}
\caption{} \label{fig:period_doubling}
\end{figure}
\end{center}

The bifurcation diagram (fig.~\ref{fig:states}) is quite complex,
and it is not possible to probe all the secondary bifurcations using
DNS.  In the present letter we focus on a narrow window ($F_0 =
4.6-4.8$) of fig. \ref{fig:states} to investigate how chaos appears
in this window.  As we will describe below, here we observe a
period-doubling route to chaos.   To understand the transition to
chaos better, we plot the phase space projections on the
($B(0,0,1)$-$B(0,0,3)$) plane.   At around $F_0 = 4.6$ we observe a
fixed point or a constant magnetic field as evident from
fig.~\ref{fig:period_doubling}(a).   At $F_0 \simeq 4.73$,  the
fixed point bifurcates to a periodic solution (limit cycle) through a Hopf
bifurcation (also shown in fig.~\ref{fig:period_doubling}(a)). The power spectral
density (PSD) plot of the time series for this $F_0$ exhibits a peak at
a single frequency $f_1 = f \simeq 0.018$ (in non-dimensional units) as
shown in the right side of   fig.~\ref{fig:period_doubling}(a).   As
we increase $F_0$, we observe  period-2, period-4, and period-8
solutions at $F_0 = 4.77, 4.777$, and 4.778 respectively.
These new states are generated through the ``period-doubling
bifurcations''.  Projections of the phase space on the
($B(0,0,1)$-$B(0,0,3)$)  plane along with the power spectra for
these three states are shown in
figs.~\ref{fig:period_doubling}(b,c,d) respectively.    The power
spectra clearly indicate peaks at subharmonics $f_2 = f/2$, $f_3 =
f/4$, $f_4 = f/8$ corresponding to the period-2, period-4, and period-8
dynamo states.  At $F_0 \simeq 4.8$, the system becomes
chaotic as evident from the phase space projection and the broad
frequency spectrum shown in fig.~\ref{fig:period_doubling}(e).  The
above observations indicate that  chaos appears at $F_0 \simeq 4.8$
through the period-doubling route to chaos.  The value of magnetic
Reynolds number for these states is around 22.  The bifurcation
diagram (fig.~\ref{fig:states}) exhibits several other windows of
chaos whose origin has not been explored in this letter.

In conclusion, our numerical simulations of dynamo for $P_M=1$  with Taylor-Green forcing reveal that the dynamo
transition takes place through a supercritical pitchfork
bifurcation. After the primary bifurcation, the system exhibits
several windows of constant, periodic, quasiperiodic and chaotic
solutions. We also find multiple coexisting attractors for a given
parameter; different initial conditions take the system to one or
the other attractor. A careful analysis of one of the chaotic
windows reveals that the dynamo becomes chaotic through a
period-doubling route to chaos.

Our numerical simulations reveal several dynamo states that have
been observed in experiments (such as VKS), earlier numerical simulations, and
low-dimensional models.  The geometry and the forcing of our
simulations however are simpler than those of experiments.  Yet the
above similarities may be due to certain inherent common features
of dynamo.  Future numerical simulations with more realistic
geometry and forcing functions will reveal valuable insights into
this puzzle.

We thank S. Fauve, E. Dormy, D. Carati, K. Kumar, and T. Lessinnes for fruitful discussions and comments. This work was supported by a research grant of DST India as Swarnajayanti fellowship to MKV.

\newpage
\begin{center}
{\bf Figure captions}\\
\end{center}

\noindent
Fig. \ref {fig:spectrum}: The  kinetic energy spectrum $\langle E^u(k)\rangle$ and the magnetic energy spectrum $\langle E^b(k)\rangle$ for $F_0=4.8$ for which the magnetic field is chaotic.  The spectra indicates that the simulations are well resolved.


\smallskip
\smallskip
\noindent Fig. \ref {fig:3D}: Flow
structure of the Taylor-Green (TG) flow:  (a) volume rendering of the velocity field magnitude that illustrates 16 TG cells due to
$k_0 = 2$;  (b) volume rendering of the magnetic field magnitude illustrating two major and two minor slabs.  This emphasizes the dominance of ${\bf B}(0,0,1)$ and ${\bf B}(0,0,2)$ magnetic modes.

\smallskip
\smallskip
\noindent Fig. \ref {fig:vector}:  The magnetic
field vector plot for a cross-section of fig.~\ref{fig:3D}(b) at $z=4.08$.

\smallskip
\smallskip
\noindent Fig. \ref {fig:time_series_states}: The
time series of the real part of ${{B}}_x(0,0,1)$ for
$F_0=3.9$(a), 10(b), 36(c) and $4.8$(d).

\smallskip
\smallskip
\noindent Fig. \ref {fig:states}: Bifurcation diagram exhibiting various dynamo
states of our simulations with FP = fixed point, P = periodic state, QP =
quasiperiodic state, and C = chaotic state. The time-averaged amplitude of
$B(0,0,1)$ is plotted for various $F_0$.  Multiple attractors are observed at
$F_0 = 16$ and $F_0 = 36$ (see
fig.~\ref{fig:mp}).  The circled region of the inset exhibits
period-doubling route to chaos (see
fig.~\ref{fig:period_doubling}).

\smallskip
\smallskip
\noindent Fig. \ref {fig:mp}:  Phase space projections
on the ($B(0,0,1)$-$B(0,0,3)$) plane for $F_0=16$(a,b) and
$F_0=36$(c,d) exhibiting multiple attractors. Figures (a,b) show coexisting fixed point and periodic states for two different
initial conditions. Figures (c,d) show coexisting
quasiperiodic and chaotic states.

\smallskip
\smallskip
\noindent Fig. \ref {fig:period_doubling}: Phase space
projections on the ($B(0,0,1)$-$B(0,0,3)$) plane and the power
spectra exhibiting period-doubling route to chaos: (a) a fixed point
at $F_0 = 4.6$, and period-1 solution at $F_0 = 4.73$;  the power
spectrum in RHS exhibits a peak at $f \simeq 0.018$ (non-dimensional units)
for $F_0 = 4.73$;  (b) period-2 solution at $F_0 = 4.77$ with power
spectrum showing two peaks at $f_1$ and $f_2 (=f/2)$; (c) period-4
solution at $F_0 = 4.777$ with power spectrum  with three peaks at
$f_1, f_2$ and $f_3 (=f/4)$; (d) period-8 solution at $F_0 = 4.778$
with power spectrum showing four peaks at $f_1, f_2, f_3$ and
$f_4 (=f/8)$; (e) chaotic solution with a broad-band power spectrum.


\newpage

\newpage

\newpage

\newpage

\newpage

\newpage

\newpage

\end{document}